\documentclass[preprint, 10pt, numbers]{sigplanconf}

\usepackage{amsmath}
\usepackage{graphicx}
\usepackage{subfigure}
\usepackage{xspace}
\usepackage{color}
\usepackage{listings}
\usepackage{courier}
\usepackage{verbatim}

\newcommand{\sysname}{{\small \sf SOSG}\xspace}

\newcommand{\para}[1]{\smallskip\noindent\textbf{#1}}

\newcommand{\mytitle}[1]{Debugging OpenStack Problems Using a State Graph Approach}

\begin{document}

\setlength{\pdfpageheight}{\paperheight}
\setlength{\pdfpagewidth}{\paperwidth}

\conferenceinfo{CONF 'yy}{Month d--d, 20yy, City, ST, Country}
\copyrightyear{20yy}
\copyrightdata{978-1-nnnn-nnnn-n/yy/mm}
\copyrightdoi{nnnnnnn.nnnnnnn}



\title{\mytitle}

\authorinfo{Yong Xiang, Hu Li, Sen Wang}
           {Tsinghua University}
           {\{xiangy13, lihu12, wangs12\} @mails.tsinghua.edu.cn}
\authorinfo{Wei Xu}
           {Tsinghua University}
           {weixu@mail.tsinghua.edu.cn}

\maketitle

\begin{abstract}

It is hard to operate and debug systems like OpenStack that integrate many independently developed modules with multiple levels of abstractions.  A major challenge is to navigate through the complex dependencies and relationships of the states in different modules or subsystems, to ensure the correctness and consistency of these states.  We present a system that captures the runtime states and events from the entire OpenStack-Ceph stack, and automatically organizes these data into a graph that we call \emph{system operation state graph} (\sysname).  With \sysname we can use intuitive graph traversal techniques to solve problems like reasoning about the state of a virtual machine.  Also, using graph-based anomaly detection, we can automatically discover hidden problems in OpenStack.  We have a scalable implementation of \sysname, and evaluate the approach on a 125-node production OpenStack cluster, finding a number of interesting problems.

\end{abstract}




\section{Introduction}

So-called cloud computing infrastructures are designed to organize servers, networking devices and storage systems into a virtual resource pool, hoping to simplify system operation.  OpenStack~\cite{openstack} is a very popular open source cloud management system.  However, even after several years of development, there are still many issues with OpenStack, making the system itself challenging to operate.

Beyond the commonly discussed code quality control issues in the OpenStack developer community, we believe systems like OpenStack are fundamentally hard to debug and operate due to the way they are designed.

OpenStack integrates many open source or proprietary software systems to perform different tasks, and the management system itself also has a number of asynchronously connected modules.  There are six major components in OpenStack, namely Nova (compute), Neutron (networking), Cinder (block storage), Swift (object storage), Glance (images) and Keystone (authentication), as well as a number of optional components to provide layer-3 routing, accounting etc.  Each module maintains its own database, and communicates through a persistent queue service~\cite{rabbitmq}.

Many of the modules provide extensible interfaces, allowing different backends for actual implementation.  For example, in our configuration,  Nova uses \emph{libvirt}~\cite{libvirt} to control a \emph{kvm}~\cite{kvm} hypervisor to provide server virtualization, and Neutron uses \emph{Open vSwitch (OVS)}~\cite{ovs} to provide virtual networks.  To make things even more complicated, it is common to use Ceph~\cite{ceph} as the storage backend, introducing another complex distributed system.


OpenStack and many similar systems are hard to operate mainly because of the following reasons.

1) The system states are distributed over the entire system and exist in many levels of abstraction, with considerable amount of duplications.  For example, the relevant states of a virtual machine (VM) exist in Nova, libvirt / kvm, Neutron, Open vSwitch, the routing agent, as well as the storage service.  For the VM to work, two conditions must hold: 1) all these components work, and 2) their states are consistent.  Unfortunately, neither of the two always holds in OpenStack.

2) Development benefits from the modularity in OpenStack design, but maintaining it requires the operator to understand all modules, an impossible task.  Each module has its own set of tools for state monitoring and other operations, and these components are located at different machines.

Even experienced operators may spend lots of time looking into the piles of manuals and running many commands to answer a simple query like \emph{should I shut down physical machine A, which VMs would be affected?} The question is harder than just listing the VMs on the machine, as the operator also needs to know what other services run on the machine.  E.g., there might be a routing service or a disk block on the server that other VMs are actively using.

3) Existing automated debugging tools, both log-based~\cite{2} and code analysis based~\cite{static-diagnosis}, only work on a single code base with consistent identifiers, there is no system-wide ID in OpenStack, especially crossing different layers and integrating with Ceph.  In addition, the state updates are asynchronous and hard to track.

In this paper, we propose a novel approach to capture the knowledge about the system runtime states in a graph, which we call System Operation State Graph (\sysname).  \sysname is designed to solve problems for operators. We show that with a simple procedure, we can automatically construct the graph, which reveals many hidden links among different system modules.  \sysname is designed to be general.  We do not assume much knowledge about the target system, but only need the log file locations and a list of commands to extract states from different modules.  Specifically, we do not need to understand the semantics of these data.  Also, all required raw data are at a component level rather than the system-level, and thus easy to provide by module developers.

Specifically, \sysname captures \emph{entities} in the entire OpenStack system at different layers.  There are many different types of entities.  VM, network and storage blocks are all entities.  For an entity, we also capture its \emph{states}, such as Nova database record and libvirt states for a VM.  We not only  keep the current state, but also previous states.  We also capture the \emph{events} (e.g. log messages) related to the entity, which are useful in debugging.

\sysname automatically discovers the links among different entities, even cross multiple modules, using a syntactic string matching on automatically discovered identifiers in events and states.  E.g. based on common strings in file / directory names and unique IDs in logs, we can infer the relationship between a VM and a Ceph data block it uses.


We present two applications of \sysname.  1) we turn ad hoc system state queries into graph traversals, simplifying these queries, especially those spanning multiple modules;  2) we perform automatic anomaly detection to find VMs that behave differently, which might indicate problems.

We implement \sysname both on Neo4j~\cite{neo4j} and GraphX~\cite{graphx} and evaluate it using real data from a professionally operated, production-quality 125-node OpenStack-Ceph cluster with over 100 active users.  We process 40 GB raw data into a 43-million-vertex, 57-million-edge graph, on which we perform traversals and anomaly detections using a GraphX cluster.  We successfully detect and diagnose many hidden or user-visible failures such as resource reclaim issues, state inconsistencies, and VM migration failures.

In summary, our contributions in this paper include:

1) We propose an approach that organizes information about system states and events into a single graph representation, \sysname, with which we can solve many complicated state queries with a common graph traversal.

2) We design an anomaly detection algorithm that automatically analyzes the state graph and find many problems based on the graph structure.

3) We provide two scalable implementations of the \sysname.  Preliminary evaluations using real operation data from a production OpenStack cluster show promising results.

%
%
%

The remaining of the paper is organized as follows: we review related work in Section 2.  We show how to process the raw system operation data into a state graph in Section 3, and present the applications of the state graph in Section 4.  We report the results of several case studies in Section 5.  And finally, we discuss the future work and conclude in Section 6 .

\section{Related Work}

\para{Automatic system diagnostics. } Bugs are inevitable in systems, and people have designed many approaches to automatically detect systems bugs, using both static analysis of the source code and runtime data such as logs.  The authors of~\cite{2} detect problems using common identifiers in text logs.  It is important to combine multiple data sources.  California Fault Lines~\cite{4} combines router configurations, syslogs with email logs to recover from failures of an email service. SherLog~\cite{static-diagnosis} is an example of using powerful static code analysis to help improving logging.  And ~\cite{LinTan} finds similar code patterns to build actionable alert prediction model. All these methods are limited to a single code base, which we do not have.

Debugging multiple frameworks is a new topic.
Pivot Tracing~\cite{1} monitors multiple frameworks in distributed systems using dynamic instrumentation, and supports relational operator to process the collected data on the fly.  Paper~\cite{3} diagnoses distributed system performance changes by tracing request flows end-to-end across components. Adding the traces creates common identifiers in a heterogeneous system.  We only use existing information in systems without extra instrumentations, and thus easier to deploy.

Anomaly detection has been widely used in system problem detection~\cite{wangc}.  \cite{16} takes advantage of Spark to perform large scale anomaly detection, and use it to detect VM performance problems.  \cite{17} uses anomaly detection to find faults in a multi-tier web system with redundancy.  These projects use a small number of homogeneous data source, while we mainly focus on analyzing states cross different system components.

\para{Anomaly detection methods. } There are many anomaly detection methods for different types of data.  ~\cite{11,12} provide techniques to simplify data from heterogeneous sources to improve anomaly detection result. Distance-based anomaly detection \cite{13,14,15, pathsim} are special techniques allowing the anomaly to be described by a probability model.  Graph anomaly detection is also a well studied topic~\cite{graph-anomaly-survey,subgraph-anomaly}.

\para{Scalable graph computation. } The recent development of efficient graph computation frameworks, such as Pregel~\cite{pregel}, Power-Graph \cite{graphlab}, GraphX and Apache Giraph \cite{giraph}, enables our approach.  Specifically, we use GraphX to process the giant state graph efficiently.

\para{Knowledge base and knowledge graphs.  } The property graph is a special case of the knowledge base (KB), a well-studied topic in data mining. There are many popular knowledge base systems, such as Knowledge Vaults \cite{knowledge-vault}, YAGO~\cite {yago1, yago2, yago3}, DBpedia \cite{dbpedia}, Freebase \cite{freebase}, and NELL \cite{nell}. And also many efforts are devoted to build these KBs \cite{kbgrow1} \cite{kbgrow2}.  Our state graph is similar to the knowledge graph, but it specifically targets machine generated system states, and thus can be built automatically.

\section{State Graph}

In this section, we introduce our core data structure, the system operation state graph, and show how we construct the graph from a number of heterogeneous raw text files.  In the next section, we introduce two applications we developed on the graph, the graph-traversal based state queries and automatic anomaly detection.

\subsection{Data structure}

Our core data structure is the state graph.  It is a special version of the property graph \cite{property-graph} automatically constructed from raw operation data.  The property graph is a directed multigraph~\cite{graphx} allowing user defined properties attached to any vertex or edge. The multigraph supports parallel edges to capture multiple relationships between the same vertex pairs.

\begin{table}[tb]
	\centering
	\caption{Data sources and their corresponding types}
	\label{tbl_operation_data}
	\begin{tabular}{ll}
		\hline
		\textbf{Type} & \textbf{Data source} \\ \hline
		DB             & OpenStack databases updates (db triggers)          \\
		Libvirt        & libvirt status \texttt{Python API}                    \\
		Ovs            & OVS status \texttt{ovsdb-client dump}              \\
		Cephimage      & Ceph image list \texttt{rbd info}                \\
		Cephfile       & Ceph block file \texttt{ls /ceph/file/dir}           \\
		\hline
		Cephlog        & log files from Ceph \\
		Log            & logs from all OpenStack components                     \\
		\hline
	\end{tabular}
\end{table}

Table~\ref{tbl_operation_data} summarizes the different data sources we use to generate the state graph.  For each \emph{record} in any data source, there is a corresponding vertex in the state graph.  We use the data source to determine the \emph{data type} of the vertex, as Table~\ref{tbl_operation_data} shows.  The vertex contains a list of key-value pairs (e.g. host:n005) as properties.

\begin{figure}
\begin{centering}
\includegraphics[width=1\linewidth]{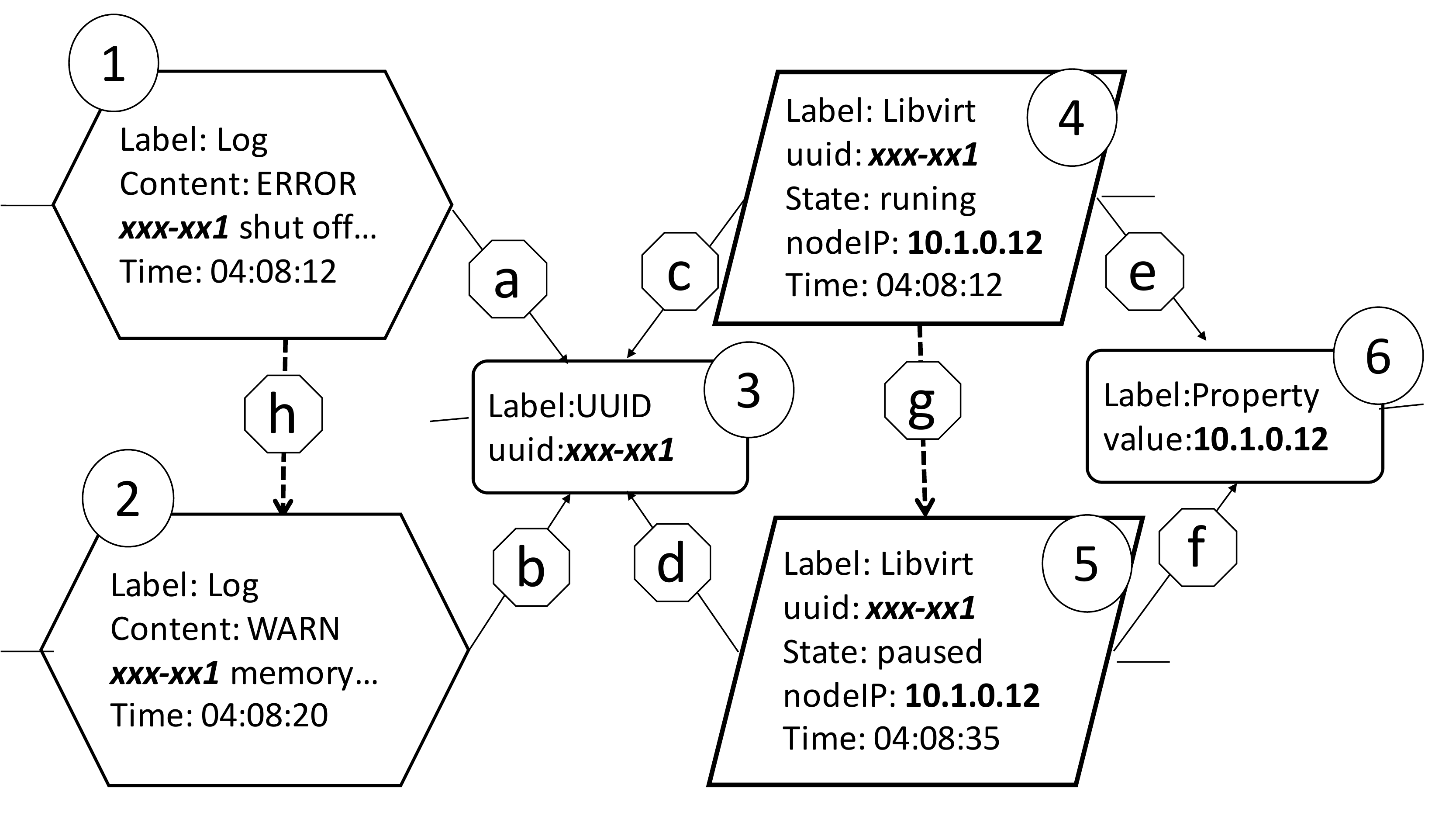}
\caption {A slice of an example state graph.  The rectangles, parallelograms, and hexagons represent entity, state and event vertices, respectively.  The numbers and letter labels are used in the discussion in the text.}
\label{fig: entity-state-event}
\end{centering}
\end{figure}

\para{Vertices. } In a state graph, there are three categories of vertices: \emph{entities}, \emph{states} and \emph{events}.  Figure~\ref{fig: entity-state-event} shows an example of a state graph with all these categories of vertices.

\emph{Entity vertices} are the centeral pieces in the state graph, as they represent instances of components or resources in the system, e.g, a VM, a disk block or a physical server. Specifically, we do not distinguish the \emph{identifier of an entity} from the \emph{entity} itself.  That is, we treat an \emph{IP address} the same as \emph{the server with the IP address}, or the \emph{UUID of a VM} the same as the \emph{VM}.  This choice is due to the limitations of the textual raw data, and the lack of need for distinction.  Vertex 3 and 6 in Figure~\ref{fig: entity-state-event} are  examples of entity vertices.

However, the raw data do not directly contain entities.  Instead, they contain \emph{states} of an entity at a specific time (e.g. a VM is running/stopped/paused in libvirt), or \emph{events} involving certain entities (e.g. a log message saying that a VM starts to shut down).   As we detail in Section~\ref{sec:graph_construction}, we extract all entity vertices from the state and event vertices in the graph.  For example, vertex 1, 2 are both event vertices and vertex 4, 5 are both state vertices in Figure~\ref{fig: entity-state-event}.

\para{Edges. } There are two types of edges in the state graph: \emph{spatial edges} and \emph{temporal edges}.  \emph{Spatial edges} capture the relationship between an entity and its states (entity-state), as well as its associated events (entity-event).  Note that in our current representation, we do not have entity-entity edges.  Instead, we represent an entity-entity relationship using a path of (entity1 $\rightarrow$ state / event $\rightarrow$ entity2).  Note that a state or event vertex acts as the ``bridge'' between two entities.  In Figure~\ref{fig: entity-state-event}, edges a, b, c, d, e, f are all spatial edges.  There is an edge between vertex 1 and 3 because the event contains the entity.  From the figure we can infer the relationship between entity vertices 3 (xxx-xx1) and 6 (10.1.0.12) by following the path (3 $\rightarrow$ d $\rightarrow$ 5 $\rightarrow$ f $\rightarrow$ 6).

\emph{Temporal edges} represent the time order of states and events that connect to the same entity.  The temporal edges make it easy to traverse events or state changes in time.  The edge always points to the increasing time direction.  Edge g and h in Figure~\ref{fig: entity-state-event} are both temporal edges.

%
%
%

\subsection{State graph construction}
\label{sec:graph_construction}

Our goal is to construct the state graph from the raw data sources in Table~\ref{tbl_operation_data}, without using any semantic information.  Here we outline our construction algorithm.

\para{Step 1: Parse raw text data to generate event and state vertices. }
The raw data are heterogeneous, including free texts, semi-structured texts and structured records. They are encoded in different formats such as JSON, csv or mysql dump files.  We provide a set of parsers for each data format to extract the information from text into key-value pairs.  Each \emph{record} (as defined by the parser) from the data source  is turned into a vertex, with the data source encoded as the data type of the vertex.  Then we add the key-value pairs to the vertex as its properties.  In Figure~\ref{fig: entity-state-event}, vertices 1, 2, 4 and 5 are generated in this step.

\para{Step 2: Discover and generate entity vertices. }
Note that not every single property of the state and event vertices represents an entity.  We need to discover the property values that might be an identifier for some entity.  While manually labeling them is feasible in a small system, we decide to automatically discover them based on statistical properties, following the identifier discovery method in~\cite{2}.  Simply speaking, we find the properties with many distinct values and each value appears at multiple places, and use them as identifiers.  In OpenStack, this method works quite well.  We take these identifiers and generate an entity vertex for each distinct value.  In Figure~\ref{fig: entity-state-event}, vertices 3 and 6 are generated in this step.

\para{Step 3: Add spatial edges. }
We generate an edge connecting a state or event vertex with an entity vertex, \emph{iff} the state or event contains the entity.  Note that for efficiency reasons, in a real implementation, we combine this step with Step 2, with some careful bookkeeping when computing of the distinct values.  We omit the details here due to space constraints.    In Figure~\ref{fig: entity-state-event}, we add edges a, b, c, d, e and f.

\para{Step 4: Add temporal edges. }
To create the temporal edges, firstly we group the state and event vertices by the entity they are associated with (a many-to-many association).  For all events / states associated with an entity, we sort them by time, and create  temporal edges according to ascending time order.  This step adds the edge g and h in Figure~\ref{fig: entity-state-event}.

\para{Summary. }
After the four steps above, we have a complete state graph capturing both temporal and spatial information.  The procedure has two advantages: 1) it is fully syntax driven, using only textual and simple statistical features, without any external semantic information or human labeling; 2) every step in the procedure contains many independent operations, and thus trivial to parallelize with a graph computation framework.  Currently we generate the graph from scratch, and we are working on algorithms to incrementally update the graph to support online monitoring applications.


%
%

\section{Applications of the State Graph}

While there are many potential applications of the state graph, we present two of them in this preliminary paper.

\subsection{System state query as graph traversal}

The direct application of the state graph is answering system state queries, so that the system operators can find all state information in a homogeneous data structure with a single method - graph traversal, instead of memorizing tons of different commands.

Queries only involving a single entity is straightforward.  We only need to find the \emph{most recent} state / event vertex connected to the entity vertex, and look up its properties.  It is  trickier to discover states involving multiple entities, such as the physical location of \emph{a specific data block in a volume of a VM}.  In this case, we need to traverse the graph through a (entity $\rightarrow$ state / event $\rightarrow$ entity $\rightarrow$ state / event $\rightarrow$ \dots) path.  We can use breadth-first-search (BFS) to find the path, and can use data type information to reduce the search space for BFS. We omit the details of the optimization due to space constraints, and instead provide a concrete example in Section~\ref{graph_traversal}.

We implement the graph traversal on both Neo4j \cite{neo4j} and GraphX, and we provide a number of convenient functions for common tasks, such as \texttt{listCephfilesForVM}, or \texttt{listVMsInSubnet}.  Note that these functions are based on a common underlying graph traversal mechanism, showing that our core techniques are general.

\subsection{Anomaly detection}

While the graph traversal can find answers to specific queries from operators, however, there are millions of states in the system, and many issues remain hidden for a long time without being noticed.  For example, we have a number of resource release failures, hidden for months in our public system (detail in Section~\ref{sec:anomaly_case}). We would like to automatically analyze the entire graph to find these hidden problems.

As a preliminary attempt, we use graph-based anomaly detection.  We choose anomaly detection because it is an unsupervised algorithm, and does not require manually labeled failure samples.  The basic assumption of anomaly detection is that most of the events and entities are normal, and any deviation from the normal case indicates problems.  This is true in our production OpenStack, where most VMs have a common set of states.  Thus, our goal is to find VMs whose states are different from their peers.

\para{Feature extraction: subgraph of a VM. } The first task is to extract the \emph{features} of a VM.  The feature captures the state of a VM.  As the normal operation of a VM depends on all components the VM uses, such as network, disk and security groups, we need to find the subgraph that roots at the VM entity vertex and also includes all its dependencies.  We can naively perform a BFS starting from the VM entity to find related entities.

One practical difficulty is that many entities, such as a subnet, are shared among different VMs.  Naive BFS will expand the subgraph through this shared subnet vertex into the resources of other VMs.  Thus, we introduce an algorithm augmenting the naive BFS with a \textit{collaborative pruning} methodology to discover shared vertices, and prevent the BFS search from passing through the shared vertices.

Intuitively, the algorithm works as follows.  We start BFS traversal from every single VM vertex.  Then on every vertex along the path, we remember a list of BFS traversals that have reached it before.  During a traversal, if we reach a vertex that has been reached by another BFS, the later BFS stops there.  In this way, we not only find subgraphs rooted at each VM, but also the information about shared resources.

\para{Distance-based anomaly detection. }  We define a distance metric between two subgraphs to capture whether the two subgraphs are similar to each other.  The goal of the distance metric is to tell a problematic VM from normal ones.

We try to capture the structure information of the subgraphs in the distance metric.  Specifically, we capture the information about \emph{triplets}~\cite{graphx}.  A triplet is two vertices along with the edge.  We encode a triplet information such as the data type of the source / destination vertex, and its location in the subgraph (i.e. the depth of the BFS traversal).  Then we define the distance between the two subgraphs using a generalized Jaccard distance \cite{wiki-jaccard}.

Finally, we apply distance-based anomaly detection to find the object with fewer than $k$ neighbors within a radius of $r$.  The parameters $k$ and $r$ directly affect our detection results. As a first attempt, we determine them empirically. As a future work, we are investigating more powerful anomaly detection techniques with more intuitive parameters.

\section{Preliminary Evaluation}

We will discuss some case studies in our preliminary evaluation in this section.  Our evaluation is based on real operation data from a 125-node production cluster that runs OpenStack (Icehouse) and Ceph. There are 5 OpenStack controller nodes, 120 compute nodes, 40 of which also doubles as Ceph storage nodes.  Each node has 12 Xeon cores, 128 GB of RAM and 10 GE network. This cluster offers computation and storage services for about 100 active users.

\para{Raw data collection.}We periodically snapshot the state of libvirt (every 60 sec), OVS (every 60 sec),  Ceph image (every 600 sec) and Ceph block file (every 3600 sec, with duplicates removed).  To capture the database states, we first dump the entire OpenStack DB right at the beginning of the experiments, and create triggers to log all database updates.  We also collect log files from all OpenStack and Ceph components. The average collected operation data is 600 MB per hour from all 125 servers. The OVS state snapshot represents about half of the data as the collector does not remove the duplicate records. Logs account for about 24\% of the data. Cephfile and Libvirt occupies 15 \% and 9\% respectively. We use a Spark \cite{spark} cluster running inside the VMs in the same OpenStack cluster to analyze these data.

%

\subsection{State graph construction performance}

We evaluate the performance of constructing state graph from the raw data as discussed above.  We build the graph using a 3-day operation data that consist of a number of text files with total 40 GB.  We construct the graph following the procedure discussed in Section~\ref{sec:graph_construction}.  We use a 15-node Spark cluster to accelerate the process.

It takes 5 minutes to parse the text files and construct the event and state vertices, and  8 minutes to expand these vertices to discover entity vertices.  Then it takes 12 minutes to add all remaining edges.  The total processing time for the 3-day worth of data takes 25 minutes, which is an acceptable cost considering the convenience this graph brings to operators.  The resulting graph contains about 43.3 million vertices and 56.6 million edges. Most of the vertices are events (log entries).

\subsection{Graph traversal example}
\label{graph_traversal}

We provide a concrete query example.  Consider the query
\textit{If physical server A encounters a hard disk failure, which VMs are affected?}  To answer it in a traditional way, the operator needs to look up many information, including which blocks are stored on the disk, which Ceph image the block belongs to, where the image is used. Each of the questions requires one or more system specific commands.

\begin{figure}
\begin{centering}
\includegraphics[width=1\linewidth]{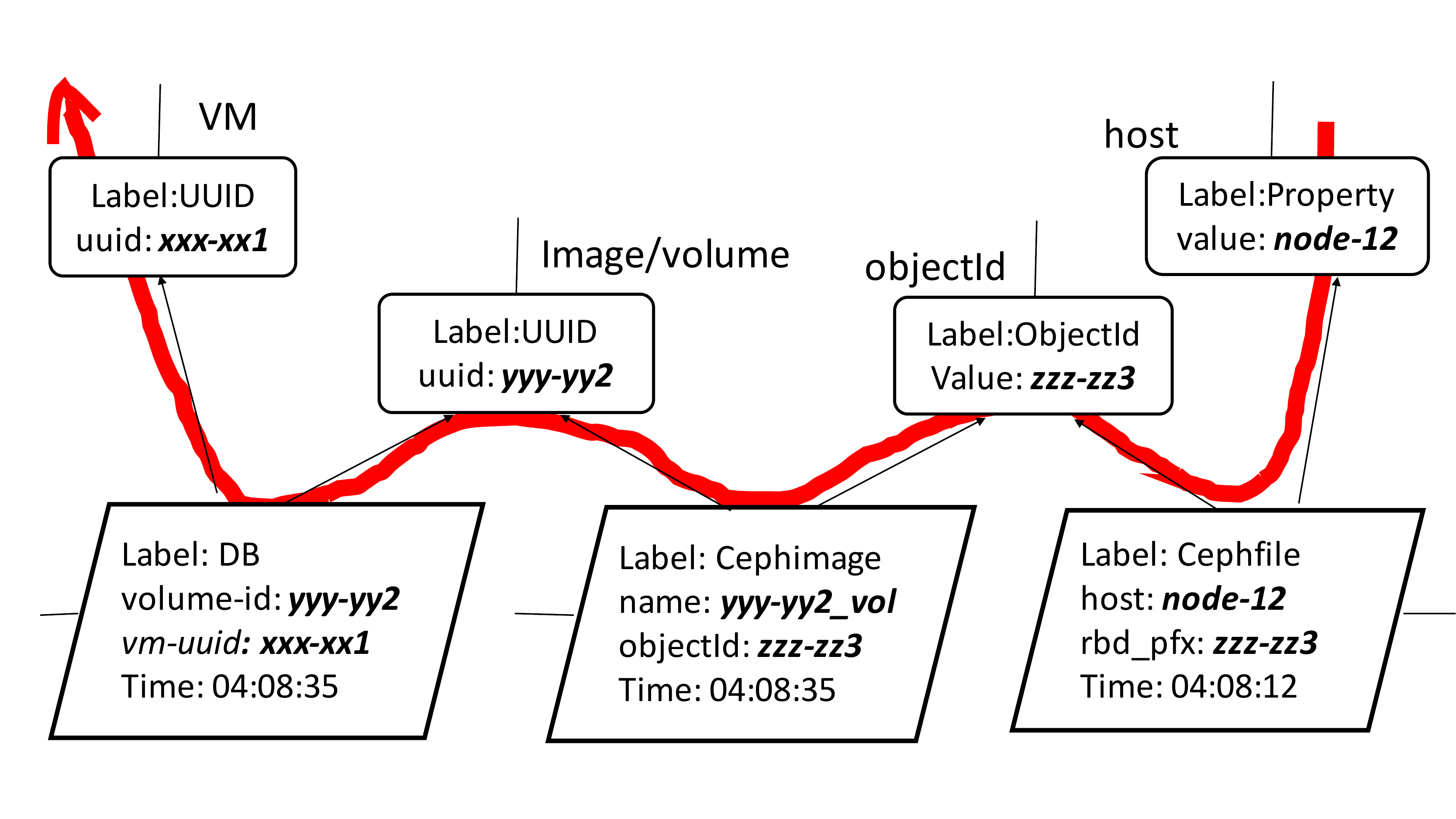}
\caption {An example path from host to VM across Ceph}
\label {fig: vm-ceph-pm}
\end{centering}
\end{figure}

Figure~\ref {fig: vm-ceph-pm} shows one of the paths that our graph traversal algorithm automatically discovers.  The path starts from a physical server and ends with a VM, acrossing the Ceph states, like (\texttt{Property} $\to$ \texttt{Cephfile} $\to$ \texttt{ObjectId} $\to$ \texttt{Cephimage} $\to$ \texttt{UUID}  $\to$ \texttt{DB} $\to$ \texttt{UUID} ). The query is fast too, and only takes 35 seconds in our setup.

\begin{figure*}[t!]
	\centering
	\subfigure[Normally deleted VM]{\label{fig: normal-case}\includegraphics[width=43mm]{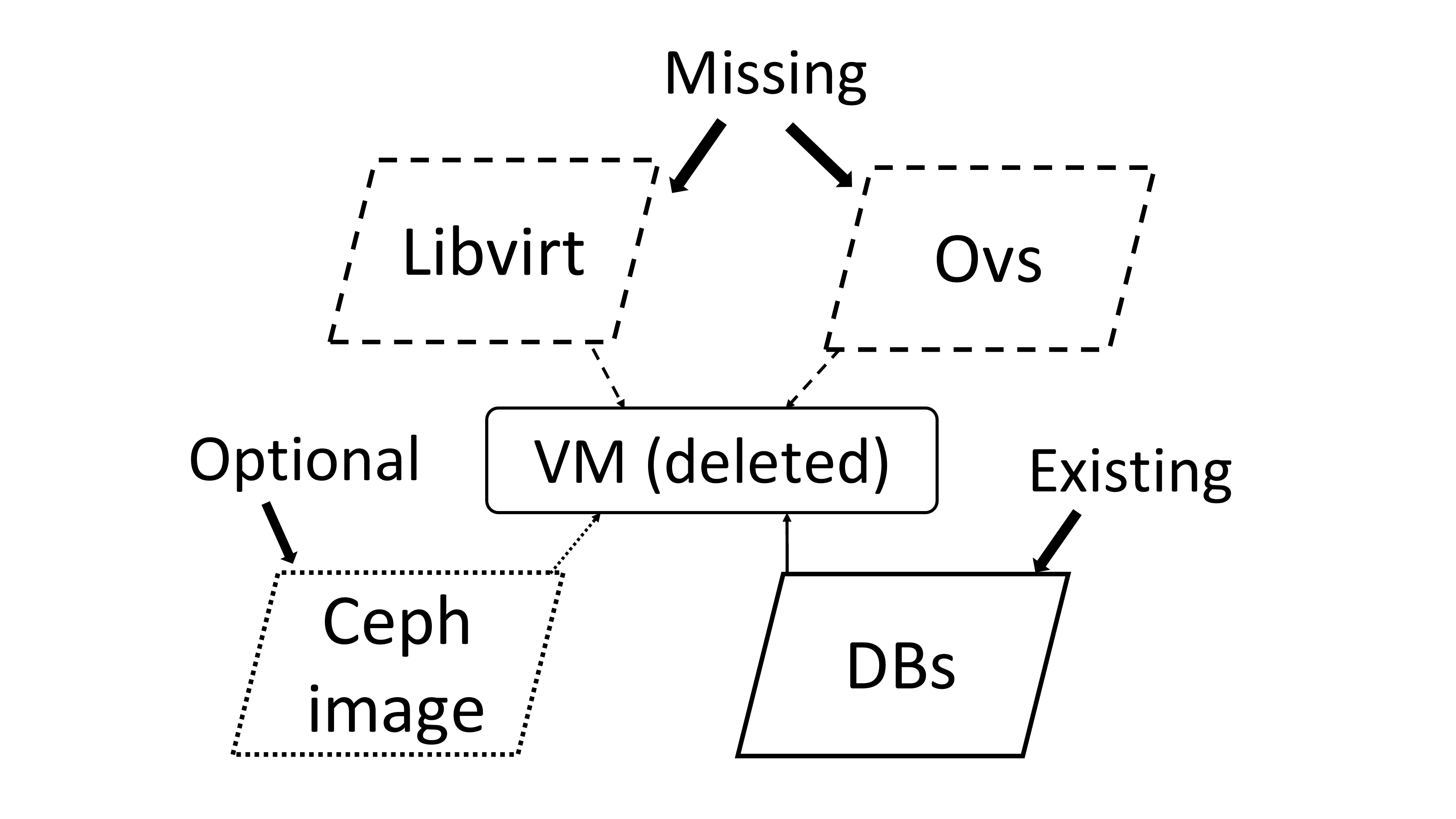}}
	\subfigure[Anomaly case 1]{\label{fig: ovs-resource-error}\includegraphics[width=43mm]{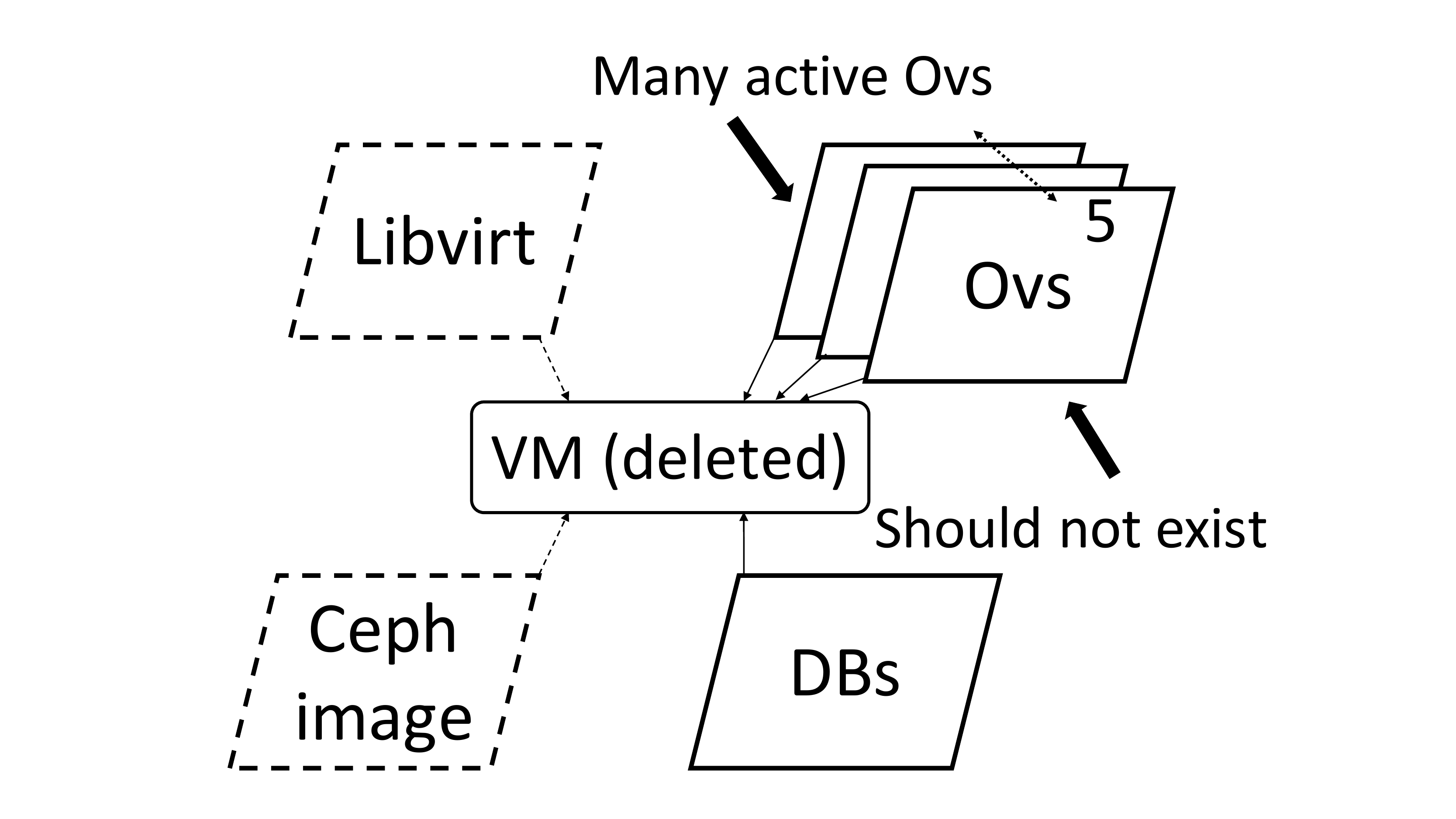}}
	\subfigure[Anomaly case 2]{\label{fig: db-mismatch-error1}\includegraphics[width=43mm]{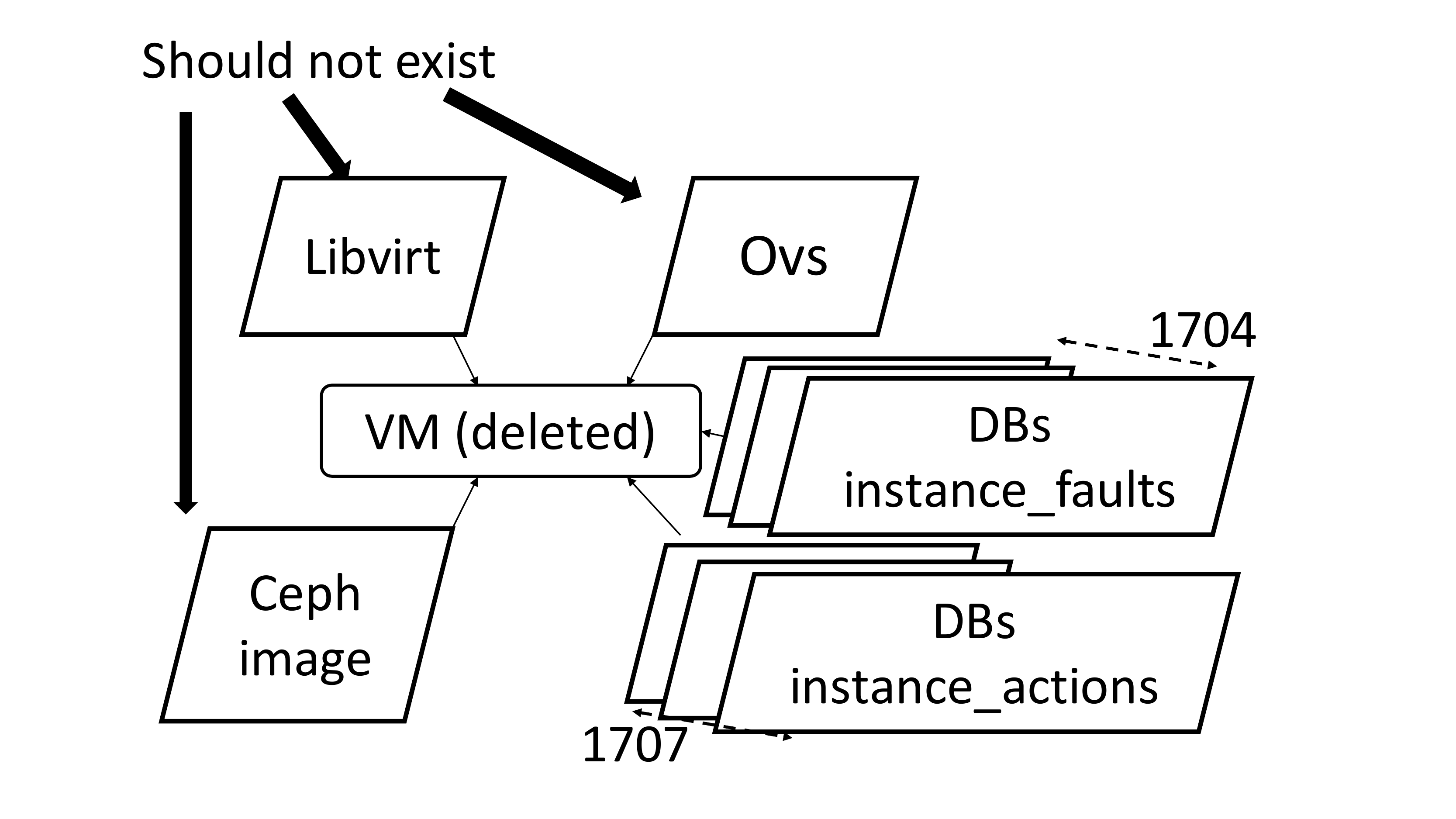}}
	\subfigure[Anomaly case 3]{\label{fig: migration-error}\includegraphics[width=43mm]{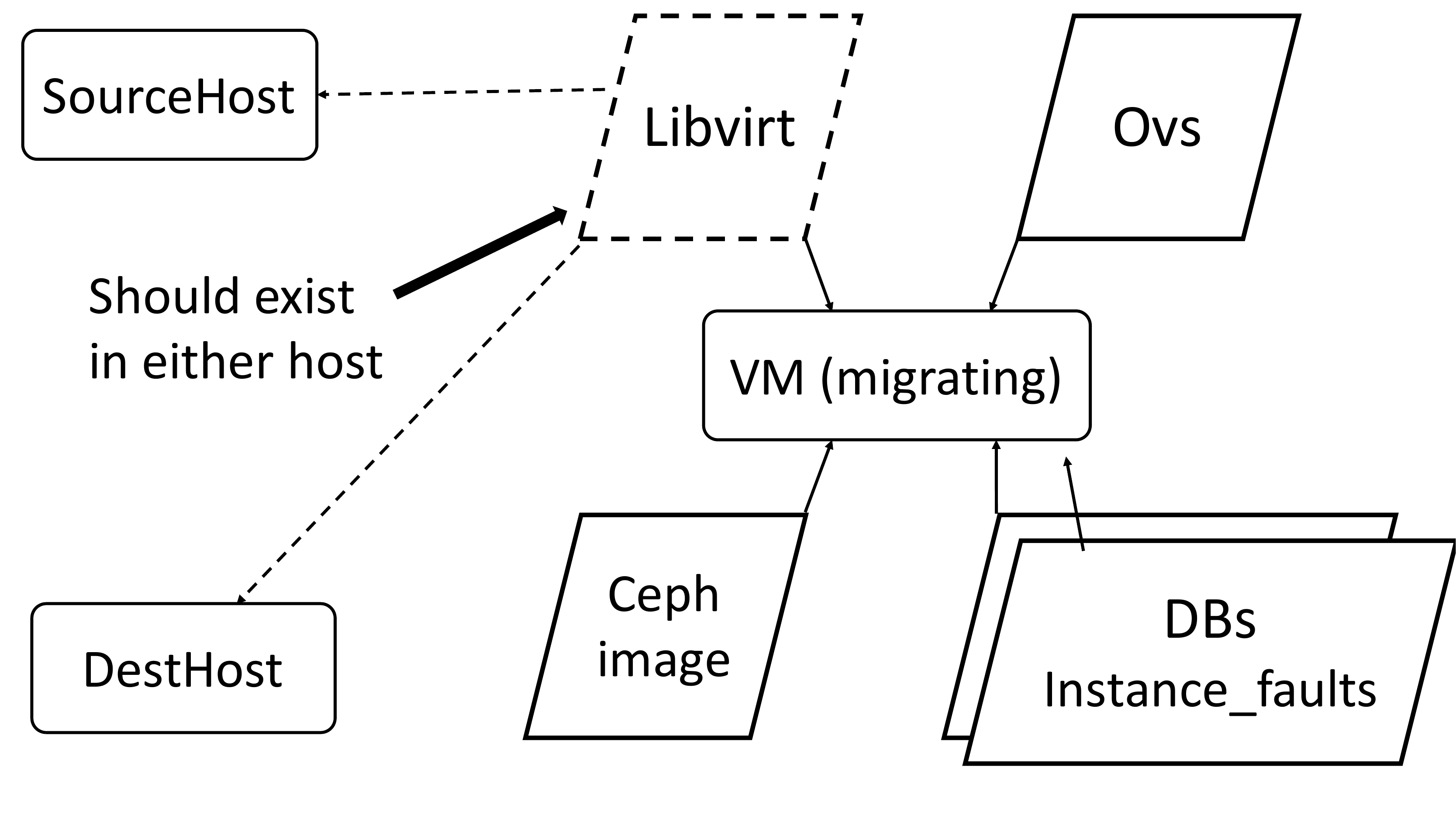}}

	\caption{Portions of normal and abnormal subgraphs.  Solid box = exist, dashed = missing, dotted lines = optional.  Same as in Figure 1, the rectangles, parallelograms and hexagons represent entity, state and event vertices, respectively.}
\end{figure*}

\subsection{Case studies: anomalies detection}
\label{sec:anomaly_case}
While we are still going through the manual process evaluating the false positive rates, we present three interesting anomalies as case studies.  The first two are both cases of hidden problems. The third one represents a complicated failure that confuses users.

\para{Case 1. Open vSwitch ports not deleted on VM deletion. } It is hard to see a resource deletion failure behind the OpenStack UI.  The anomaly detection algorithm captures a subgraph, with relevant portion shown in Figure~\ref{fig: ovs-resource-error}.  It is abnormal because the entity vertex is connected to many Ovs state vertices, whereas the normal case (Figure \ref{fig: normal-case}) does not.

As we try to confirm the graph structure-based detection is semantically valid, we inspect the properties of each vertex.  We find that the VM instance has been deleted months ago, but still has many active OVS states associated with it.  This anomaly indicates that the virtual ports of the VM are not deleted, resulting a resource leak.  Note that it is not a common problem (otherwise it will not be an anomaly). 



\para{Case 2. Database record does not match physical states. }  It is common to see disagreement of the OpenStack database record and the actual physical state.  Figure \ref {fig: db-mismatch-error1} shows a case.  The subgraph is picked out by anomaly detection mainly because there are thousands of DB state vertices directly connect to the VM entity vertex.

Looking for the semantic reason, we find that the VM has been in \texttt{deleted} state for months, but the libvirt, Cephimage and OVS states still remain.  Again, looking into the state vertices connected to the VM vertex, we can see the possible reason:  \texttt{nova.instance\_faults} shows $1,704$ failures in $1,707$ of \texttt{nova.instance\_actions}. This finding not only indicates an inconsistency case, but also suggests some serious bugs in OpenStack's retrying / recovery mechanism.


\para{Case 3. Failed VM Migration. }  Lastly, we present a more complex and user visible failure during the VM migration.  The user reported the issue as a freezing migration process.  The user has nothing to do but delete the VM.  Anomaly detection algorithm picks out the abnormal subgraph too, as Figure~\ref{fig: migration-error} shows.  The subgraph is abnormal in that the migrating VM is missing libvirt state, both from the source host and the destination host.

A closer manual inspection shows that during this migration from node-118 to node-38 exception happened , and two database state vertices \texttt{nova.instance\_faults} connected to the VM show that \texttt{cannot remove config /etc/libvirt/qemu/instance-0000155e.xml: Read-only file system} on the source and \texttt{error removing image} on the destination.   As a result, the storage (virtual disk) of the VM migrated but the computation did not, causing a failed migration. In addition, the missing libvirt state vertices on both node-118 and node-38 serve as another evidence of the unsuccessful migration.

An even deeper inspection at the logs and events associated with this VM vertex indicates that the VM encountered a migration problem: there are 653 repeated \texttt{Instance not resizing, skipping migration} records out of all 1653 log lines. This repeated skipping of a small-instance (2 VCPU, 4 GB RAM) migration also suggests some bugs in OpenStack's resource management.


\section{Conclusion and Future Work}

We will focus on the following important directions as our future work.

\para{Root cause analysis with events and state history. }  As we indicate (with manual analysis) in Section~\ref{sec:anomaly_case}, event sequences (logs) are an indication \emph{why} the system ends up in an inconsistent state.  We would like to have a model that maps logs to the corresponding anomalous state.  The model might help predict the failure before it actually happens.

\para{Including other data sources. }  The state graph is a great way to automatically discover links among different information about a system, both runtime and static.  We want to incorporate other static data sources, such as the source code, bug reports and documentations into the graph, and hope to provide more insights into how to \emph{fix} the bugs discovered.

\para{Applying \sysname to other systems. }  Though we have only presented \sysname as a tool for debugging OpenStack, we believe the approach is general.  We would like to apply it to detect problems in other distributed systems, such as big data frameworks and general web services organized in a service oriented architecture (SOA).

\para{Conclusion. }As both researchers and system operation practitioners, we keep wondering \emph{``What is the core set of knowledge in system operation?"}  Most of the times, we believe it is the experience of knowing about all dependencies, or links, among different system components, and the knowledge about different tools to inspect and change the states of these components.  Many of the knowledge is too trivial to remember, impossible to transfer to a new system, and hard to teach to another person.  All of these problems make system operation hard.

The above is our motivation to build \sysname, which captures the runtime information, including both states and events, and discover the hidden links among these pieces of information.  By leveraging modern graph computation capacity, we can process a vast amount of redundant data and automatically construct the graph.  With the graph, we turn the typical task such as ad hoc probing of different system components into an intuitive graph traversal problem, making the exploration of heterogeneous systems easier.  We also develop a subgraph-based anomaly detection method to automatically analyze system states to find hidden problems.  We evaluate \sysname with data from our production OpenStack cluster with tens of components, and demonstrate its effectiveness.


\bibliographystyle{abbrvnat}


\end{document}